\documentclass[a4paper, 12pt]{article}
\usepackage{amsmath, amssymb, graphics}
\newcommand{\mathsym}[1]{{}}
\newcommand{\unicode}[1]{{}}

\catcode`\@=11

\@addtoreset{equation}{section} 

\def\dddot#1{\mathinner{\buildrel\vbox{\kern5pt\hbox{...}}\over{#1}}}
\def\ddddot#1{\mathinner{\buildrel\vbox{\kern5pt\hbox{....}}\over{#1}}}

\begin{document}

\begin {center}
{\Large Symmetries and Integrability of Modified Camassa-Holm Equation with an Arbitrary Parameter}\\[3 mm]
{\small A Durga Devi $ ^ 1$, K Krishnakumar$ ^ {2*}$ , R Sinuvasan$ ^ 3$ and PGL Leach$ ^ 4$\\[3 mm]
$^ 1 $ Department of Physics, Srinivasa Ramanujan Centre, SASTRA Deemed to be University, Kumbakonam 612 001, India.\\$^ 2 $ Department of Mathematics, Srinivasa Ramanujan Centre, SASTRA Deemed to be University, Kumbakonam 612 001, India.\\$^ 3 $ Department of Mathematics, SASTRA Deemed to be University, Thanjavur 613 401, India.\\$^4 $Department of Mathematics, Durban University of Technology, PO Box 1334,\\ Durban ~4000,~ Republic of South Africa.}
\end{center}
\begin {abstract}
We study the symmetry and integrability of a modified Camassa-Holm Equation (MCH), with an arbitrary parameter $k,$ of the form $$u_{t}+k(u-u_{xx})^2u_{x}-u_{xxt}+(u^{2}-{u_{x}}^2)(u_{x}-u_{xxx})=0.$$ By using Lie point symmetries we reduce the order of the above equation and also we obtain interesting novel solutions for the reduced ordinary differential equations. Finally we apply the Painlev\'e Test to the resultant nonlinear ordinary differential equation.
\end {abstract}
{\bf MSC Numbers:}  17B80; 34M15; 58J70\\
{\bf PACS Numbers:} 02.20.Sv; 02.30.Hq; 02.30.Ik\\
{\bf Keywords:} Modified Camassa-Holm Equation (MCH), Symmetry, Painlev\'e Test, Integrability.

\section {Introduction}
The linear and nonlinear partial differential equations have encompassed the landscape of many disciplines such as chemical, physical, biological sciences, engineering and mathematics due to its potential ability to provide adequate understanding of the underlying systems.
There are many wave equations in engineering and physics that are quite helpful and relevant such as the Korteweg–de Vries (KdV) equation, nonlinear coupled KdV equations, Modified Camassa-Holm (MCH) equation and so on \cite{Korteweg 95 a, Hirota 71 a, Hirota 73 a, Hirota 73 b, Hirota 76 a, Hirota 74 a, Gui-2013, Fuch-1996, Fokas1995, Xia-2018, Fu-2013}. In general, the closed- form solutions of the differential equations are very hard to achieve in most instances.

Numerous systematic techniques still have been developed to find the solutions of differential equations in recent years. Especially, Lie Symmetry Analysis \cite{Lie74a, Olver-86, Leach 88 a, Leach 99, Leach 12 a, Andriopoulos09a, Tamizhmani 14 a, Krishnakumar 14 b, Winternitz09a}, Painlev\'e Analysis \cite{Ablowitz 80a, Ablowitz 80b, Ramani 89, Feix 97 a, Andriopoulos 06 a, Lemmer 93 a}, Adomian decomposition method \cite{G. Adomian88, Wazwaz}, the variational iteration method, the homotopy analysis method,  the homotopy perturbation method and variational homotopy perturbation method have been utilized successfully to solve and study the nature of integrability of partial  differential equations \cite{Majeed}. Among these methods the symmetry analysis and Painlev\'e Analysis have been used by researchers effectively \cite{Winternitz09a, Winternitz02, Winternitz09b}.

\strut\hfill

Recently researchers have been inspired to study the integrability and solutions of various Camassa-Holm (CH) types of equations with cubic or higher-order nonlinearities. Olver and Rosenau \cite{Olver-1996} and Fuchssteiner \cite{Fuch-1996} derived the well known modified Camassa-Holm (MCH) equation:
\begin{eqnarray}
y_{t}+u y_{x}+2u_{x}y=0,\nonumber\\
y=u-u_{xx}
\label{mch}
\end{eqnarray}
 by employing the triHamiltonian duality approach to the biHamiltonian representation of the MKDV equation. The functions $u$ and $y$ represent, the velocity of the fluid and its potential density \cite{Qiao-2006}. The MCH equation is a water-wave equation and is a suitable approximation of the incompressible irrotational Euler system. Consequently Qiao \cite{Qiao-2006} discussed the nature of integrability and the structure of solutions of the Modified Camassa Holm equation.  In some literature the modified Camassa Holm equation is otherwise state as FORQ equation \cite{Himonas-2014, Yang-2017}.

 \strut\hfill

	The MCH equation is completely Integrable \cite{Olver-1996}.  It has a biHamiltonian structure and also admits a Lax pair \cite{Qiao-2011} and hence may be solved by the inverse scattering transform method. The MCH equation can be viewed as a cubic extension of the well-known Camassa-Holm (CH) equation which was proposed as a model describing the unidirectional propagation of shallow
water waves \cite{Camassaholm, ConstantinLannes, FuchssteinerFokas} and axially symmetric waves in hyperelastic rods \cite{ConstantinStrauss}.

\strut\hfill

	The purpose of this work is to study the symmetry and integrability of MCH equation with an arbitrary parameter which is given by
 \begin{eqnarray}
u_{t}+k(u-u_{xx})^2u_{x}-u_{xxt}+(u^{2}-{u_{x}}^2)(u_{x}-u_{xxx})=0,
\label{mcha2}
\end{eqnarray}
 where parameter $k \in R$ characterizes the magnitude of the linear dispersion.

 \strut\hfill

 We split this article into two sections. In the first section we discuss Lie point symmetries and reductions of order for (\ref{mcha2}). Lie symmetry is an important tool to provide a systematic method for finding the solution of the given differential equations. In which, we reduce the order of the ordinary or partial differential equations by using the symmetry. Hence, Lie symmetries plays a major role in it. Lie symmetries can be used for the classification of differential equations. Throughout this literature the Mathematica add-on Sym \cite{Dimas05a, Dimas06a, Dimas08a} is used to compute the symmetries. In the second section, we study the Painlev\'e Analysis for equation (\ref{mcha2}) and we able to prove that the equation (\ref{mcha2}) is integrable.

\strut\hfill

\section{The Symmetries of modified Camassa-Holm equation}
Qiao \cite{Qiao-2006} discussed the equation
\begin{eqnarray}
u_{t}+k(u-u_{xx})^2u_{x}-u_{xxt}+(u^{2}-{u_{x}}^2)(u_{x}-u_{xxx})=0,
\label{mcha}
\end{eqnarray}
which is integrable only when $k=2$. The author also suggested to examine the integrability of the above equation for all possible values of $k$. Therefore we propose a more general form of (\ref{mcha}) by taking $k$ as $k(t)$ which is a function of $t$. Hence the modified Camassa-Holm equation (MCH) with arbitrary function, $k(t)$, \cite{Qiao-2006, GaoLiuQu} can be represented as
\begin{eqnarray}
u_{t}+k(t)(u-u_{xx})^2u_{x}-u_{xxt}+(u^{2}-{u_{x}}^2)(u_{x}-u_{xxx})=0,
\label{mchaa}
\end{eqnarray}
when $k(t)$ is an arbitrary function of $t$. Equation $(\ref{mchaa})$ has a symmetry $\partial_x$. Also we have arrived at the trivial solution $u(x,t)=C$ to the equation $(\ref{mchaa})$.
To find the nontrivial solutions, we choose $k(t)=k$ as an arbitrary parameter. Therefore the equation becomes
 \begin{eqnarray}
u_{t}+k(u-u_{xx})^2u_{x}-u_{xxt}+(u^{2}-{u_{x}}^2)(u_{x}-u_{xxx})=0.
\label{mch}
\end{eqnarray}

The symmetries of equation(\ref{mch}), when $k$ is an arbitrary parameter, are
\begin{eqnarray}
\textbf{X}_{1}&=&\partial_{x}\\
\textbf{X}_{2}&=&\partial_{t}\\
\textbf{X}_{3}&=&2t\partial_{t}-u\partial_{u}.
\end{eqnarray}
The Lie Brackets of the Lie symmetries of equation (\ref{mch}) are\\
\begin{center}
\begin{tabular}{|c|ccc|}
$\left[ \textbf{X}_{I},\textbf{X}_{J}\right] $ & $\textbf{X}_{1}$ & $\textbf{X}_{2}$ & $\textbf{X}_{3}$  \\
$\textbf{X}_{1}$ & $0$ & $0$ & $0$   \\
$\textbf{X}_{2}$ & $0$ & $0$ & $2\textbf{X}_{2}$ \\
$\textbf{X}_{3}$ & $0$ & $-2\textbf{X}_{2}$ & $0$. \\
\end{tabular}
\end{center}
Now we discuss the travelling-wave solution of equation (\ref{mch}) by taking the linear combination of $\textbf{X}_{1}$ and $\textbf{X}_{2}$.  Therefore we can represent the linear combination of these two symmetries as $\textbf{X}_{4}=\partial_{t}+C \partial_{x}$.  The corresponding canonical variables are $r=x-Ct$ and $u(x,t)=Q(r)$. Based on these canonical variables the equation($\ref{mch})$ can be reduced to
\begin{eqnarray}
 Q^{\prime\prime\prime} (C-Q^2+{Q^{\prime}}^2)+k(Q^{\prime\prime}
 -2Q)Q^{\prime}Q^{\prime\prime}-{Q^{\prime}}^3-(C-Q^2-kQ^2)Q^{\prime}=0, \label{red2}
\end{eqnarray}
where  $Q$ is the function of the new independent variable $r$.
The symmetry of the above equation(\ref{red2}) is $\partial_{r}$ for which the canonical variables are $Q(r)=Q$ and $Q^{\prime}=W(Q)$.  Based on these canonical variables the equation, (\ref{red2}), can be reduced to
\begin{eqnarray}
(C-Q^2+W^2)WW^{\prime\prime}+(C-Q^2+W^2+kW^2){W^{\prime}}^2
-2kQWW^{\prime}&&\nonumber\\-W^2+(k+1)Q^2-C=0,\label{redsecord}
\end{eqnarray}
where $W$ is a function of $Q$.  When we find the symmetries of the equation, ({\ref{redsecord}}), we arrive the following two possible cases.\\

\strut\hfill

{\bf Case:1}

 If $k\neq-2$, Then the symmetries of the equation, ({\ref{redsecord}}), are
\begin{eqnarray}
\Gamma_{1}&=&\partial_{Q}+\frac{Q}{W}\partial_{W}\\
\Gamma_{2}&=&\frac{(C-Q^{2}+W^{2})^{\frac{-k}{2}}}{W}\partial_{W}\\
\Gamma_{3}&=&\frac{Q(C-Q^{2}+W^{2})^{\frac{-k}{2}}}{W}\partial_{W}\\
\Gamma_{4}&=&\frac{(C-Q^{2}+W^{2})}{(k+2)W}\partial_{W}\\
\Gamma_{5}&=&2{Q}\partial_{Q}+\frac{(C+(3+2k)Q^2+W^2)}{(k+2)W}\partial_{W}\\
\Gamma_{6}&=&{Q}^{2}\partial_{Q}+\frac{Q(C+(k+1)Q^2+W^2)}{(k+2)W}\partial_{W}\\
\Gamma_{7}&=&\frac{(C-Q^2+W^{2})^{\frac{k+2}{2}}}{k+2}\partial_{Q}+\frac{Q(C-Q^2+W^{2})^{\frac{k+2}{2}}}{(k+2)W}\partial_{W}\\
\Gamma_{8}&=&\frac{Q(C-Q^2+W^{2})^{\frac{k+2}{2}}}{k+2}\partial_{Q}+
\frac{(C-Q^2+W^{2})^{\frac{k+2}{2}}(C+(k+1)Q^2+W^2)}{(k+2)^2W}{\partial_{W}}.\nonumber\\
\end{eqnarray}

When we use $\Gamma_{1}$, the solution of equation, ({\ref{redsecord}}), is given by
\begin{equation}
W=\pm \sqrt{Q^2+2I_{1}},
\end{equation}
where $I_{1}$ is the constant of integration.
From $\Gamma_{4}$, the solution is
\begin{equation}
W=\pm \sqrt{(Q^2-C)}.
\end{equation}

By using $\Gamma_{5}$, the solution is
\begin{equation}
W=\pm\sqrt{Q^2-C+(2Q(k+2))^{\frac{1}{k+2}}I_{2}},
\end{equation}
where $I_2$ is the constant of integration, with $(2Q(k+2))^{\frac{2}{k+2}} I_{2}^2\bigg( Q^2-C+(2Q(k+2))^{\frac{1}{k+2}}I_{2}\bigg)=0.$

By using $\Gamma_{6}$, the solution is
\begin{equation}
W=\pm\sqrt{Q^2-C+(Q(k+2))^{\frac{2}{k+2}}I_{3}},
\end{equation}
where $I_{3}$ is the constant of integration.

\strut\hfill

{\bf Case:2}

If $k=-2$, then the symmetries of the equation ({\ref{redsecord}}) are
\begin{eqnarray}
\Gamma_{1}&=&\partial_{Q}+\frac{Q}{W}\partial_{W}\\
\Gamma_{2}&=&\frac{(C-Q^{2}+W^{2})}{W}\partial_{W}\\
\Gamma_{3}&=&\frac{Q(C-Q^{2}+W^{2})}{W}\partial_{W}\\
\Gamma_{4}&=& \log[C-Q^2+W^2]\partial_{Q}+\frac{Q\log[C-Q^2+W^2]}{W}\partial_{W}\\
\Gamma_{5}&=&\frac{(C-Q^2+W^2)\log[C-Q^2+W^2]}{W}\partial_{W}\\
\Gamma_{6}&=&4Q\partial_{Q}+\frac{4Q^2+(C-Q^2+W^2)\log[C-Q^2+W^2]}{W}\partial_{W}\\
\Gamma_{7}&=&2Q^2\partial_{Q}+\frac{2Q^3+Q(C-Q^2+W^2)\log[C-Q^2+W^2]}{W}\partial_{W}\\
\Gamma_{8}&=&2Q\log[C-Q^2+W^2]\partial_{Q}\nonumber\\&+&\frac{\log[C-Q^2+W^2](2Q^2+(C-Q^2+W^2)\log[C-Q^2+W^2])}{W}\partial_{W}.
\nonumber\\
\end{eqnarray}

When we use $\Gamma_{1}$, the solution of equation, ({\ref{redsecord}}), is given by
\begin{equation}
W=\pm \sqrt{Q^2+2I_{4}}.
\end{equation}

From $\Gamma_{2}$, the solution is
\begin{equation}
W=\pm \sqrt{(Q^2-C)}.
\end{equation}

From $\Gamma_{3}$, the solution is
\begin{equation}
W=\pm \sqrt{Q^2-C}.
\end{equation}

By using $\Gamma_{4}$ along with the solution obtained from $\Gamma_{1}$, the another solution of $\Gamma_{4}$ is
\begin{equation}
W=\pm\sqrt{1-C+Q^2}.
\end{equation}

By using $\Gamma_{5}$, the solution is
\begin{equation}
W=\pm\sqrt{1-C+Q^2}.
\end{equation}

By using $\Gamma_{6}$, the solution is
\begin{equation}
W=\pm\sqrt{Q^2-C+e^{\pm i{\sqrt{\frac{Q}{I_{5}}}}}}
\end{equation}
with condition $\bigg(e^{ i{\sqrt{\frac{Q}{I_{5}}}}}(C-Q^2)-1\bigg)((k+2)\sqrt{Q}-2i\sqrt{I_{5}})$=$0.$

By using $\Gamma_{7}$, the solution is
\begin{equation}
W=\pm\sqrt{Q^2-C+e^{-\frac{Q}{I_{6}}}}
\end{equation}
with condition $(k+2)\bigg((C-Q^{2})e^{\frac{Q}{I_{6}}}-1\bigg)$=$0.$

By using $\Gamma_{8}$, the solution is
\begin{equation}
W=\pm\sqrt{Q^2-C+e^{-\frac{Q}{I_{7}}}}.
\end{equation}
In the above $I_{4}, I_5, I_6$ and $I_7$ are the constants of integration.

\strut\hfill

In what follows, we reduce the order of another form of the equation (\ref{mch}) by using an equivalence form of the symmetry, $\textbf{X}_{3}$, as follows. Therefore, firstly consider the another form of equation (\ref{mch}),
\begin{eqnarray}
v_{t}+k(t)v^{2}u_{x}+(u^{2}-u_{x}^{2})v_{x}=0,\nonumber\\
v=u-u_{xx}.
\label{Amch}
\end{eqnarray}
 According to equation, $(\ref{Amch})$, the equivalence form of $\textbf{X}_{3}$ is represented as ${\bf\Gamma}_{3}=2t\partial_{t}-u\partial_{u}-v\partial_{v}$. Also the associated transformations for the variables are: $u(x,t)=\frac{y(x)}{\sqrt{t}}$ and $v(x,t)=\frac{z(x)}{\sqrt{t}}$. Therefore the equation (\ref{Amch}) becomes
\begin{eqnarray}
2(y^2-{y^{\prime}}^2)z^{\prime}+2k z^2 y^{\prime}-z=0\nonumber\\
y^{\prime\prime}-y+z=0.\label{SE}
\end{eqnarray}
 The symmetry of the above system $(\ref{SE})$ is $\partial_x$. The reduced ordinary differential system from system $(\ref{SE})$ is given by
 \begin{eqnarray}
2(y^2-p^2)q+2k z^2 p-z=0,\nonumber\\
p p^{\prime}-y+z=0,\label{FE}
\end{eqnarray}
by using the canonical variables $y(x)=y$, $z(x)=z$, $y^{\prime}=p(y)$ and $z^{\prime}=q(z)$ which are obtained from the symmetry $\partial_x$. One can find the solution of (\ref{Amch}) by solving the equation (\ref{FE}).

\section{Painlev\'e Analysis}


Paul Painlev\'e, was French mathematician, was found the Painlev\'e Test. By applying this test one can determine the integrability of a given ordinary differential equations or partial differential equations in terms of analytic functions \cite{Ramani 89, Painleve97, Painleve00, Painleve02}.

\strut\hfill

Mark Feix and his collaborators have investigated the procedure of Ablowitz Ramani Segur (ARS) Algorithm also they found the underlying relation between ARS algorithm and Painlev\'e test \cite{Ablowitz 80a, Ablowitz 80b, Ramani 89}. Based on their outcomes, the solution can be identified as two possible ways(i.e) Right Painleve Series and Left Painleve Series.  The Right painleve series means the Laurent expansion in an ascending form obtained from negative integral power and it constitutes the solution of the given differential equation within a punctured disc centred on the singularity \cite{Andriopoulos 06 a, Krishnakumar 16}. On the other hand, the left painleve series means an expansion descending from the same negative integral power constitutes the solution of the given differential equation shows without the disc \cite{Andriopoulos 06 a, Krishnakumar 16}.

\strut\hfill

The existence of nongeneric negative and  positive resonances was demonstrated by Andriopoulos and Leach in \cite{Andriopoulos 06 a}. They proved that  the solution of the given differential equation in annulus can be studied by standard singularity analysis.

\strut\hfill

In this paper, we can also identify the given differential equation (\ref{mcha}) is integrable, when the ordinary differential equation (\ref{red2}) which obtained from (\ref{mcha}) posses Painlev\'e Property. Also we can find the corresponding solution to the differential equation (\ref{red2}).

 \subsection{Painlev\'e Test}

 Consider the equation (\ref{mcha})
\begin{equation}
u_{t}+k(u-u_{xx})^2u_{x}-u_{xxt}+(u^{2}-{u_{x}}^2)(u_{x}-u_{xxx})=0.
\label{mcha1}
\end{equation}
 As discussed earlier, the above equation (\ref{mcha1}) has been changed to a third-order nonlinear autonomous ordinary differential equation by substituting $r=x-C t$,

\begin{eqnarray}
 Q^{\prime\prime\prime} (C-Q^2+{Q^{\prime}}^2)+k(Q^{\prime\prime}-2Q)Q^{\prime}Q^{\prime\prime}-{Q^{\prime}}^3-(C-Q^2-kQ^2)Q^{\prime}=0, \label{red22}
\end{eqnarray}
where $Q$ is a function of $r$. If we use  this form of (\ref{red22}), it does not pass the Painlev\'e Test. Therefore, we  substitute $Q(r)=\frac{1}{v[r]}$.   Then the equation becomes
 \begin{eqnarray}\label{red21}
 v^2(v^2-C v^4-{v^{\prime}}^2)v^{\prime\prime\prime}+v v^{\prime}v^{\prime\prime}(6C v^4-2(k+3)v^2+2(2k+3){v^{\prime}}^2-k v v^{\prime\prime})&&\nonumber\\-v^{\prime}(2(2k+3){v^{\prime}}^4-C v^6-(4k+7)v^2{v^{\prime}}^2+v^4(1+k+6C {v^{\prime}}^2))=0.
 \end{eqnarray}
 According to the Painlev\'e Test the dominant terms of (\ref{red21}) are given by
\begin{equation}\label{ceq}
v^2(-C v^4-{v^{\prime}}^2)v^{\prime\prime\prime}+v v^{\prime}v^{\prime\prime}(6C v^4+2(2k+3){v^{\prime}}^2-k v v^{\prime\prime})-6C v^4{v^{\prime}}^3-2(2k+3){v^{\prime}}^5
\end{equation}
The behavior of leading-order is $a_{-1} w^{-1}$, where $w=(x-x_0)$.  Next, to find the resonances $(s)$ we substitute $v(w)=a_{-1} w^{-1} + m w^{-1 + s}$ into (\ref{ceq}) and equate the coefficients of $m$ to zero then we have $s=-1, \, 0 ~\text{and} ~1$.

\strut\hfill

 Now we are substituting the expansion $v(w)= - \frac{1}{w} + a_0 + a_1 w $ into (\ref{red21}) to calculate the arbitrariness of $a_0$ and $a_1$. For which collect the coefficients of various  powers of $w$ then equate them to zero. Finally, we get that $a_0~ \text{and} ~a_1$ are all arbitrary constants. Therefore, we conclude that the reduced form of the modified Camassa-Holm Equation (\ref{red21}) passes the Painlev\'{e} Test. As a consequence, the equation (\ref{mcha1}) is obviously an integrable partial differential equation.

\section {Conclusion}

We have examined the Lie point symmetries and Painlev\'e Test for the modified Camassa-Holm Equation with an arbitrary parameter. By the performance of a symmetry analysis we have successfully found the Lie point symmetries of equation (\ref{mcha1}) with the arbitrary parameter $k$. Consequently we have reduced the order of the equation (\ref{mcha1}) from third-order to second-order and then we have also obtained a set of solutions for the resultant second-order ordinary differential equation (\ref{redsecord}). Through the Painlev\'e Test, we were observed that the MCH equation (\ref{mcha1}) is an integrable partial differential equation even with an arbitrary parameter $k$.

\section*{Acknowledgements}
 ADD and KK thank Prof. Stylianos Dimas, S\'ao Jos\'e dos Campos/SP, Brasil, for providing us a new version of the SYM-Package.

\end{document}